\documentclass[twocolumn, twocolappendix]{aastex701}

\usepackage{multirow} 
\usepackage{threeparttable}
\usepackage{amsmath} 

\begin{document}

\title{Origin of the Fe K$\alpha$ red excess  in IC 4329A with XRISM and multi-wavelength spectroscopy}

\author[orcid=0000-0003-2161-0361]{Misaki Mizumoto}
\affiliation{Science Education Research Unit, University of Teacher Education Fukuoka, Munakata, Fukuoka 811-4192, Japan}
\email[show]{mizumoto-m@fukuoka-edu.ac.jp}  

\author[orcid=0009-0000-9577-8701]{Yuya Nakatani}
\affiliation{Department of Astronomy, Kyoto University, Kitashirakawa-Oiwake-cho, Sakyo-ku, Kyoto 606-8502, Japan
}
\email{nakatani@kusastro.kyoto-u.ac.jp}

\author[orcid=0000-0002-9754-3081]{Satoshi Yamada}
\affiliation{The Frontier Research Institute for Interdisciplinary Sciences, Tohoku University, Aramaki, Aoba-ku, Sendai, Miyagi 980-8578, Japan
}

\affiliation{Department of Astronomy, University of Geneva, Ch.d’Ecogia 16, 1290, Versoix, Switzerland}
\affiliation{Astronomical Institute, Tohoku University, Aramaki, Aoba-ku, Sendai, Miyagi 980-8578, Japan
}
\email{satoshi.yamada@terra.astr.tohoku.ac.jp}

\author[orcid=0000-0001-7821-6715]{Yoshihiro Ueda}
\affiliation{Department of Astronomy, Kyoto University, Kitashirakawa-Oiwake-cho, Sakyo-ku, Kyoto 606-8502, Japan
}
\email{ueda@kusastro.kyoto-u.ac.jp}

\author[orcid=0000-0002-6808-2052]{Taiki Kawamuro}
\affiliation{Department of Earth and Space Science, The University of Osaka, 1-1 Machikaneyama, Toyonaka 560-0043, Osaka, Japan
}
\email{kawamuro@ess.sci.osaka-u.ac.jp}

\author[orcid=0000-0002-5701-0811]{Shoji Ogawa}
\affiliation{Faculty of Science and Technology, Tokyo University of Science, 
2641 Yamazaki, Noda-shi, Chiba-ken 278-8510, Japan
}
\email{ogawa.shoji@rs.tus.ac.jp}

\author[orcid=0000-0003-3224-1821]{Yuto Mochizuki}
\affiliation{Institute of Space and Astronautical Science (ISAS), Japan Aerospace Exploration Agency (JAXA), 3-1-1 Yoshinodai, Chuo-ku, Sagamihara, Kanagawa 252-5210, Japan}

\affiliation{Department of Astronomy, Graduate School of Science, The University of Tokyo, 7-3-1 Hongo, Bunkyo-ku, Tokyo 113-0033, Japan
}
\email{mochizuki@ac.jaxa.jp}

\author[orcid=0009-0006-4377-4219]{Kanta Fujiwara}
\affiliation{Department of Astronomy, Kyoto University, Kitashirakawa-Oiwake-cho, Sakyo-ku, Kyoto 606-8502, Japan
}
\email{fujiwara@kusastro.kyoto-u.ac.jp}

\author{Fumihide Iwamuro}
\affiliation{Department of Astronomy, Kyoto University, Kitashirakawa-Oiwake-cho, Sakyo-ku, Kyoto 606-8502, Japan
}
\email{iwamuro@kusastro.kyoto-u.ac.jp}

\author{Shogo Otsubo}
\affiliation{Laboratory of Infrared High-resolution spectroscopy (LiH), Koyama Astronomical Observatory, Kyoto Sangyo University, Motoyama, Kamigamo, Kita-ku, Kyoto, 603-8555, Japan}
\email{otsubo@cc.kyoto-su.ac.jp }

\author[orcid=0009-0003-6827-6396]{Noriyuki Matsunaga}
\affiliation{Department of Astronomy, Graduate School of Science, The University of Tokyo, 7-3-1 Hongo, Bunkyo-ku, Tokyo 113-0033, Japan
}
\affiliation{Laboratory of Infrared High-resolution spectroscopy (LiH), Koyama Astronomical Observatory, Kyoto Sangyo University, Motoyama, Kamigamo, Kita-ku, Kyoto, 603-8555, Japan}
\email{matsunaga@astron.s.u-tokyo.ac.jp}

\author[orcid=0000-0002-2861-4069]{Daisuke Taniguchi}
\affiliation{Department of Physics, Tokyo Metropolitan University, 1-1 Minami-Osawa, Hachioji, Tokyo 192-0397, Japan
}
\affiliation{Laboratory of Infrared High-resolution spectroscopy (LiH), Koyama Astronomical Observatory, Kyoto Sangyo University, Motoyama, Kamigamo, Kita-ku, Kyoto, 603-8555, Japan}
\email{d.taniguchi.astro@gmail.com }

\author{Hideyo Kawakita}
\affiliation{Laboratory of Infrared High-resolution spectroscopy (LiH), Koyama Astronomical Observatory, Kyoto Sangyo University, Motoyama, Kamigamo, Kita-ku, Kyoto, 603-8555, Japan}
\affiliation{Department of Astrophysics and Atmospheric Sciences, Faculty of Science, Kyoto Sangyo University, Motoyama, Kamigamo, Kita-ku,
Kyoto 603-8555, Japan
}
\email{kawakthd@cc.kyoto-su.ac.jp }

\begin{abstract}
We present high-resolution X-ray spectroscopy of the bright local Seyfert~1
galaxy IC~4329A obtained with \textit{XRISM}/Resolve, supported by
simultaneous \textit{NuSTAR} data and quasi-simultaneous optical/near-infrared
(NIR) spectroscopy. Resolve separates the Fe K$\alpha$ line into a narrow core
and a broad component exhibiting a red excess. We examined three scenarios for
this red excess: (1) a redshifted X-ray broad line region (xBLR),
(2) relativistic disk reflection, and (3) an unshifted xBLR partially absorbed
by a mildly ionized absorber. All three models reproduce the X-ray spectra
equally well. Our optical (H$\beta$) and NIR (Pa$\beta$) spectroscopy shows
that the broad lines are redshifted by
$v \sim +640\text{--}1050\text{ km s}^{-1}$, close to the solution of the
redshifted xBLR model
($v_{\rm shift} = +1700_{-800}^{+1700}\text{ km s}^{-1}$). Motivated by this,
we constructed a hybrid model in which the xBLR shares the velocity structure
of the optical/NIR BLR and the remaining red excess is attributed to a
relativistic disk line, which also provides an acceptable fit. The inferred
xBLR radius is $R_{\rm BLR} \sim (3\text{--}6) \times 10^3\ R_g$
($\sim 12\text{--}22$ light-days).
From the narrow Fe K$\alpha$ core in the hybrid model
(${\rm FWHM} = 870_{-280}^{+260}\text{ km s}^{-1}$), we estimate a torus
distance of $R_{\rm torus} \sim 0.9_{-0.4}^{+1.0}$~pc
($\sim 1100$ light-days), close to the dust sublimation radius.
Furthermore, two ultra-fast outflow (UFO) absorption features are identified with $v_{\text{wind}} \sim -0.30c$ and $-0.24c$, both confirmed at $>99\%$ confidence.

\end{abstract}



\section{Introduction} 
Active Galactic Nuclei (AGNs) comprise a complex, multi-scale architecture including an accretion disk, a broad line region (BLR), and a dusty molecular torus, as conceptualized in the unified model \citep{1995PASP..107..803U}. In the X-ray regime, the $\mathrm{Fe~K\alpha}$ fluorescent line at 6.4~keV serves as an indispensable diagnostic of the geometry and physical conditions of the circumnuclear material surrounding the central supermassive black hole \citep[SMBH; e.g.,][]{1991MNRAS.249..352G, 2007MNRAS.382..194N}. In particular, the narrow component ($\mathrm{FWHM} \lesssim 2000~\mathrm{km~s^{-1}}$) is generally attributed to fluorescence from distant, cold gas \citep[e.g.,][]{2024ApJ...973L..25X}. Accurately determining the spatial location and physical nature of this Fe-K reprocessor is fundamental to mapping the circumnuclear structure of AGNs.

To constrain the reprocessor scale, spectroscopic line-profile analysis provides a direct and powerful method. The Resolve microcalorimeter \citep{ishisaki2025, kelley2025} onboard the X-Ray Imaging and Spectroscopic Mission (\textit{XRISM}; \citealt{XRISM}) has achieved an energy resolution better than $5~\mathrm{eV}$ across the $\mathrm{Fe~K}$ band, corresponding to a velocity resolution of a few hundred $\mathrm{km~s^{-1}}$. This unprecedented spectral precision allows us to dissect the complex profile of the $\mathrm{Fe~K\alpha}$ emission line and unambiguously disentangle distinct physical components, such as cold reflection from the distant torus and broader fluorescence arising from the BLR or photoionized winds \citep[e.g.,][]{2024ApJ...973L..25X, 2025PASJ...77S.209B, 2025ApJ...994L..13K, 2026A&A...706A.255L, 2026ApJ..1003....6F}. Consequently, \textit{XRISM}/Resolve enables simultaneous and self-consistent mapping of both the outer circumnuclear structure and the inner sub-parsec accretion dynamics.

IC~4329A ($z=0.01605$; \citealt{1991AJ....101...57W}) is one of the X-ray brightest Seyfert 1 galaxies in the local Universe. Due to its minimal line-of-sight intrinsic absorption and prominent, narrow $\mathrm{Fe~K\alpha}$ emission, it offers an ideal laboratory for probing the unabsorbed continuum and its reprocessed radiation from the surrounding structures. Based on fundamental plane studies \citep{2001MNRAS.327..265H, 2002AJ....123.2990B}, its distance is estimated to be $D = 59 \pm 9~\mathrm{Mpc}$ \citep{2009AJ....138..323T}. Optical reverberation mapping constraints yield a black hole mass of $M_{\mathrm{BH}} = 6.8_{-1.1}^{+1.2} \times 10^7 M_{\odot}$, a BLR inclination angle of $\theta_{i,\mathrm{BLR}} = 40_{-18}^{+27}{}^\circ$, and an opening angle of $\theta_{o,\mathrm{BLR}} = 69_{-30}^{+15}{}^\circ$ \citep{bentz2023}.
{Its bolometric luminosity is about $\sim1\times10^{45}$~erg~s$^{-1}$ \citep{2022ApJS..261....2K}, which corresponds to $\sim10$\% of the Eddington luminosity.}

Despite extensive study, the precise radial location of the $\mathrm{Fe~K}$ reprocessor in IC~4329A remains debated. X-ray reverberation mapping attempts have yielded conflicting results, largely due to differences in temporal sampling and observation epochs. \citet{2016ApJ...821...15F} reported significant $\mathrm{Fe~K\alpha}$ flux variations ($\sim 10\%$--$20\%$) on timescales of $\sim 2000~\mathrm{days}$, suggesting a reprocessor location at several light-years. Conversely, \citet{2022A&A...664A..46A} identified short-term line variability on timescales of days, placing an upper limit of $\lesssim 100~\mathrm{light\text{-}days}$ on the reprocessor radius, which points toward the outer BLR or the inner edge of the torus. High-resolution spectroscopy with \textit{XRISM}/Resolve is therefore essential to resolve this long-standing ambiguity.

In this paper, we present high-resolution X-ray spectroscopy of IC~4329A obtained with \textit{XRISM}/Resolve. To securely constrain the broadband continuum and high-energy reflection features (e.g., the Compton hump), simultaneous hard X-ray observations were conducted with \textit{NuSTAR}. Furthermore, to investigate the dynamical link between X-ray and other bands, we performed quasi-simultaneous optical and near-infrared (NIR) high-resolution spectroscopy using Seimei/KOOLS-IFU \citep{2020PASJ...72...48K,2019PASJ...71..102M,2025PASJ...77.1065M} and Magellan/WINERED \citep{2022PASP..134a5004I,2024SPIE13096E..31O}. 

The structure of this paper is organized as follows. In Section~\ref{sec:obs}, we detail the X-ray, optical, and NIR observations and their respective data reduction procedures. Section~\ref{sec:results} presents the spectral modeling of the X-ray $\mathrm{Fe~K}$ complex alongside the optical/NIR emission line profiles. In Section~\ref{sec:discussion}, we discuss the physical origins of the redshifted xBLR, the narrow $\mathrm{Fe~K\alpha}$ line, and the detected ultra-fast outflow (UFO) features. Finally, our conclusions are summarized in Section~\ref{sec:conclusion}. Throughout this paper, statistical uncertainties are quoted at the $90\%$ confidence level unless stated otherwise.

\section{Observations and data reduction} \label{sec:obs}

\subsection{\textit{XRISM} and \textit{NuSTAR}}

\textit{XRISM} observations were carried out from 2025-08-07 to 11, accumulating a net exposure time of $146~\mathrm{ks}$ (OBSID: 201044010). During this observation, the Resolve instrument \citep{kelley2025,ishisaki2025} was operated in the OPEN filter wheel position with the gate valve closed.
A portion of the \textit{XRISM} observation was conducted simultaneously with \textit{NuSTAR} \citep{nustar}, yielding a net exposure time of $26~\mathrm{ks}$ (OBSID: 91101617002).

Data reduction and spectral analysis were performed using HEASoft v6.36 \citep{heasoft,ftools}. 
Although data processing generally followed standard pipelines, standard processing partially failed due to telemetry anomalies encountered during the observation. 
We therefore performed customized data reduction and reprocessed the Resolve event data following instructions provided by the \textit{XRISM} Science Data Center\footnote{Details are available in the \textit{XRISM} operation log page (\url{https://xrism.isas.jaxa.jp/research/observers/operation_log/Satellite/index.html}).}.
The data were processed using pre-pipeline version \texttt{005\_004.28Feb2025\_V002.015}, pipeline script version \texttt{03.02.014.010}, and CALDB version \texttt{20260315}. 
Only High-resolution Primary (Hp) events were selected in the Resolve spectroscopy. 
The Response Matrix File (RMF) was generated using the extra-large (``X'') option to fully incorporate instrumental effects. 
The response files were created based on the cleaned event file with Low-resolution Secondary (Ls) events excluded, in order to avoid potential contamination from pseudo-events\footnote{\url{https://heasarc.gsfc.nasa.gov/docs/xrism/analysis/abc_guide/xrism_abc.pdf}}. 
The 2.4--12~keV energy band was used for the spectral fitting.
Throughout this paper, spectral fitting was performed using optimal binning scheme \citep{2016A&A...587A.151K} alongside $C$-statistics \citep{cash1979}, although the visual binning in certain figures has been modified for illustrative clarity.
The Non-X-ray Background (NXB) for Resolve was estimated using the NXB database and modeled accordingly\footnote{\url{https://heasarc.gsfc.nasa.gov/docs/xrism/analysis/nxb/index.html}}. Both the source and NXB models were fitted simultaneously to the observed spectrum. For visual clarity, NXB-subtracted spectra are presented in all spectral figures in this work.

For the \textit{NuSTAR} data analysis, source photons were extracted from a circular region with a radius of $50^{\prime\prime}$, while the background was extracted from a surrounding annulus with inner and outer radii of $150^{\prime\prime}$ and $250^{\prime\prime}$, respectively.
The CALDB version \texttt{20250812} was used.
Above $35~\mathrm{keV}$, the \textit{NuSTAR} spectrum becomes dominated by background; hence, higher-energy data were omitted from our analysis.

To assess the goodness of fit, we computed the expected value ($C_{\rm e}$)
and variance ($C_{\rm v}$) of the $C$-statistic for each best-fit model
following \citet{Kaastra2017}. Instead of the approximate formulae given
there, we evaluated $C_{\rm e}$ and $C_{\rm v}$ exactly by summing the 
Poisson distribution of the predicted model counts in each spectral bin.
A fit is regarded acceptable when the observed $C$-statistic is within
a few $\sigma$ ($\equiv \sqrt{C_{\rm v}}$) of $C_{\rm e}$. Strictly, this
method applies to Poisson data without background subtraction, which holds
for the Resolve spectrum because its NXB is modeled rather than subtracted.
For the \textit{NuSTAR} spectra, for which background spectra are used
(i.e., effectively the $W$-statistic), we applied the same calculation as
an approximation; this is justified because the background contributes
only $\lesssim$1\% of the total counts in the 3--35 keV band.

\subsection{Seimei/KOOLS-IFU}
Optical integral-field spectroscopy was carried out on 2025 April 26, approximately three months prior to the \textit{XRISM} campaign, using the 3.8~m Seimei Telescope \citep{2020PASJ...72...48K} equipped with KOOLS-IFU \citep{2019PASJ...71..102M, 2025PASJ...77.1065M} under program IDs 25A-N-CT01 and 25A-K-0012. KOOLS-IFU utilizes a 110-fiber array providing an $8.4'' \times 8.0''$ field of view, corresponding to a physical scale of approximately $2.4 \times 2.3~\mathrm{kpc}$ at $D = 59~\mathrm{Mpc}$. We employed the VPH495 grism covering $4300$--$5900~\text{\AA}$ with a spectral resolving power of $R \approx 1500$, corresponding to a velocity resolution ($\mathrm{FWHM}$) of $\sim 200~\mathrm{km~s^{-1}}$.
The total on-source exposure time for IC~4329A was $900~\mathrm{s}$ ($300~\mathrm{s} \times 3$).

Data reduction was conducted using \textsc{IRAF} \citep{1986SPIE..627..733T, 1993ASPC...52..173T} and the standard KOOLS-IFU pipeline. Basic calibration included bias subtraction, flat-fielding, cosmic-ray removal, and spatial fiber-distortion correction using arc lamp and flat-field frames. Sky background was subtracted using adjacent sky exposures of equal duration. One-dimensional spectra were extracted by co-adding the spectra
from fibers within a $6^{\prime\prime}$-diameter aperture centered on the centroid,
using equal weights. Absolute flux calibration was established using the spectrophotometric standard star HR~5501 (observed with an exposure of $5~\mathrm{s} \times 10$). 
For visual clarity in presentation, the spectrum was lightly smoothed to suppress high-frequency noise, although all spectral fitting was performed on the unsmoothed data.

\subsection{Magellan/WINERED}
Near-infrared (NIR) high-resolution spectroscopy was obtained using the Warm INfrared Echelle spectrograph to Realize Extreme Dispersion and sensitivity \citep[WINERED;][]{2022PASP..134a5004I,2024SPIE13096E..31O} mounted on the 6.5~m Clay/Magellan II telescope at Las Campanas Observatory. WINERED covers $9100$--$13500~\text{\AA}$. We adopted the WIDE mode with a $0.60^{\prime\prime}$ slit width ($4.0~\mathrm{pixels}$), yielding a resolving power of $R = 18000$ ($\Delta v = 17~\mathrm{km~s^{-1}}$).
Observations were carried out on 2025 August 14, three days after the \textit{XRISM} campaign.
Target exposures for IC~4329A consisted of $60~\mathrm{s} \times 4$ and $120~\mathrm{s} \times 2$ (total exposure of $480~\mathrm{s}$).
Sky background was subtracted using off-source sky frames taken with identical exposure times.
Data reduction was executed using the automated WARP pipeline \citep[][]{2024PASP..136a4504H}.
Our analysis focuses on the $12000$--$13200~\text{\AA}$ region (orders 42 to 46), encompassing the $\mathrm{Pa}\beta$ line.

Telluric absorption features in orders 43--46 were corrected using \textsc{TerraPCA}\footnote{\url{https://github.com/nmatsuna/TerraPCA}} \citep{2026ApJS..284...49M}, a Principal Component Analysis framework that models complex telluric variations using standard star observations with intrinsic stellar lines masked out. 
Because order 42 lies outside the spectral coverage of \textsc{TerraPCA}, its telluric features were corrected using the \texttt{telluric} task in \textsc{IRAF} with the A-type standard star HD~160365 (exposure $2~\mathrm{s} \times 4$). The spectra were normalized per order, telluric-corrected, restored to their flux scale, stitched together, and normalized across orders 42--46 with a linear continuum fit. The stitched spectrum was smoothed for presentation purposes only (Figure~\ref{fig:opticalBLR}b).

\section{Results} \label{sec:results}

\subsection{Optical/NIR spectra} \label{sec:result_optical}
First we study the optical H$\beta$ and NIR Pa$\beta$ broad lines. 
We performed spectral decomposition of the optical Seimei/KOOLS-IFU spectrum using \textsc{pyQSOfit} \citep{2018ascl.soft09008G, 2019ApJS..241...34S, 2024ApJ...974..153R}. The rest-frame spectrum was decomposed into host-galaxy light and AGN continuum via PCA. After subtracting the host galaxy, AGN continuum, and optical \ion{Fe}{2} empirical template, the emission-line complex around $\mathrm{H}\beta$ and [\ion{O}{3}]~$\lambda\lambda 4959, 5007$ was fitted with multiple Gaussian components: three Gaussians for $\mathrm{H}\beta$ and two Gaussians for each [\ion{O}{3}] line. 

\begin{figure}[tb]
   \centering
    \includegraphics[width=\columnwidth]{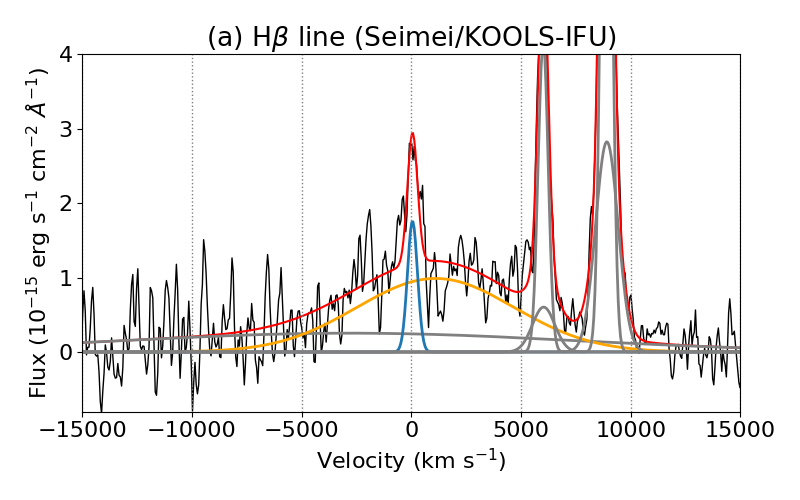}
    \includegraphics[width=\columnwidth]{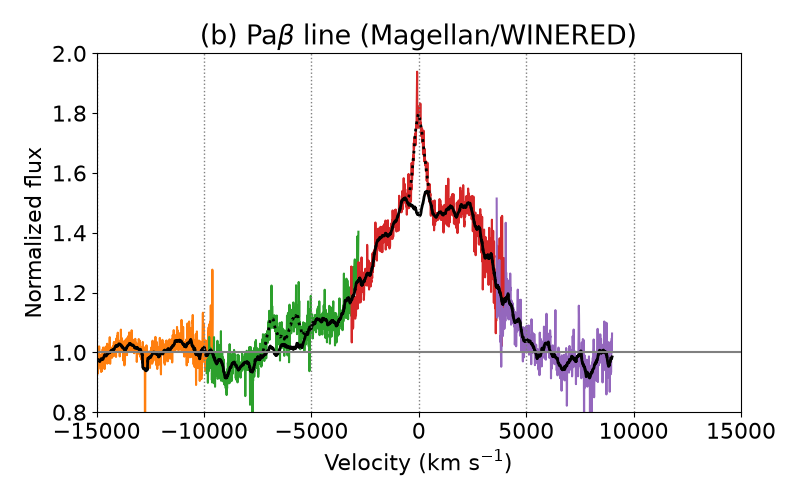}
            \caption{
            (a) Optical $\mathrm{H}\beta$ spectral region observed with Seimei/KOOLS-IFU. Black data points show the observed spectrum. Red, blue, and orange lines correspond to the total best-fit model, the narrow H$\beta$ component, and the redshifted broad H$\beta$ component, respectively. Gray lines represent other decomposed components (e.g., host galaxy, AGN continuum, and [\ion{O}{3}] lines). (b) High-resolution NIR spectrum around $\mathrm{Pa}\beta$ obtained with Magellan/WINERED. Colored data points display individual echelle orders. The solid black line represents the stitched and smoothed spectrum after subtracting narrow emission lines ($\mathrm{Pa}\beta$, [\ion{Fe}{2}] $\lambda 12570.2$, and [\ion{S}{9}] $\lambda 12523.0$), which are indicated by dashed lines.
            }
         \label{fig:opticalBLR}
   \end{figure}

Figure~\ref{fig:opticalBLR}(a) illustrates the velocity profile, where zero velocity ($0~\mathrm{km~s^{-1}}$) is defined by the peak of the narrow $\mathrm{H}\beta$ component. The broad $\mathrm{H}\beta$ emission component exhibits a redshift of $+1050 \pm 460~\mathrm{km~s^{-1}}$, with the line broadening of $\mathrm{FWHM}=9700\pm1000$~km~s$^{-1}$.

Figure~\ref{fig:opticalBLR}(b) presents the NIR Magellan/WINERED spectrum focusing on $\mathrm{Pa}\beta$.
To measure the broad-line centroid velocity $v_0$, we evaluated profile symmetry by comparing the observed profile $D_1(v)$ within $|v| < 4000~\mathrm{km~s^{-1}}$ against its velocity-reversed profile $D_2(v) \equiv D_1(2v_0 - v)$.
The optimal velocity shift $v_0$ was determined by maximizing the Pearson correlation coefficient between $D_1(v)$ and $D_2(v)$, with uncertainties ($90\%$ confidence level) estimated via Monte Carlo simulations.
This analysis yields a broad $\mathrm{Pa}\beta$ redshift of $v_0 = +640 \pm 30~\mathrm{km~s^{-1}}$. The line broadening is $\mathrm{FWHM}\sim6000$~km~s$^{-1}$.

In summary, both the optical and NIR datasets consistently reveal a velocity offset of approximately $600\text{--}1000~\mathrm{km~s^{-1}}$ in the BLR.
While the estimated FWHM values differ between the two lines, literature values have been reported as $6472~\mathrm{km~s^{-1}}$ \citep[observed on 2004-05-21;][]{2017ApJ...850...74K} and $5906 \pm 656~\mathrm{km~s^{-1}}$ \citep[observed on 2012-04-16;][]{2022ApJS..261....5M}.
This comparison implies that the FWHM derived from the Seimei data is likely biased toward higher values, possibly because the relatively low $\mathrm{S/N}$ of the spectrum prevents the extended broad component from being tightly constrained.

%This redshifted broad-line signature is consistently present across different epochs and wavelength regimes.
%Figure~\ref{fig:opticalBLR}(b) shows the optical spectrum extracted from BASS DR2 \citep{2022ApJS..261....2K}, obtained on 2012 April 16 with ESO-VLT/X-shooter \citep{2011A&A...536A.105V}. Applying the same decomposition procedure yields a broad $\mathrm{H}\beta$ velocity shift of $+740 \pm 70~\mathrm{km~s^{-1}}$.

\subsection{X-ray spectra}
\begin{figure}[tb]
   \centering
    \includegraphics[width=\columnwidth]{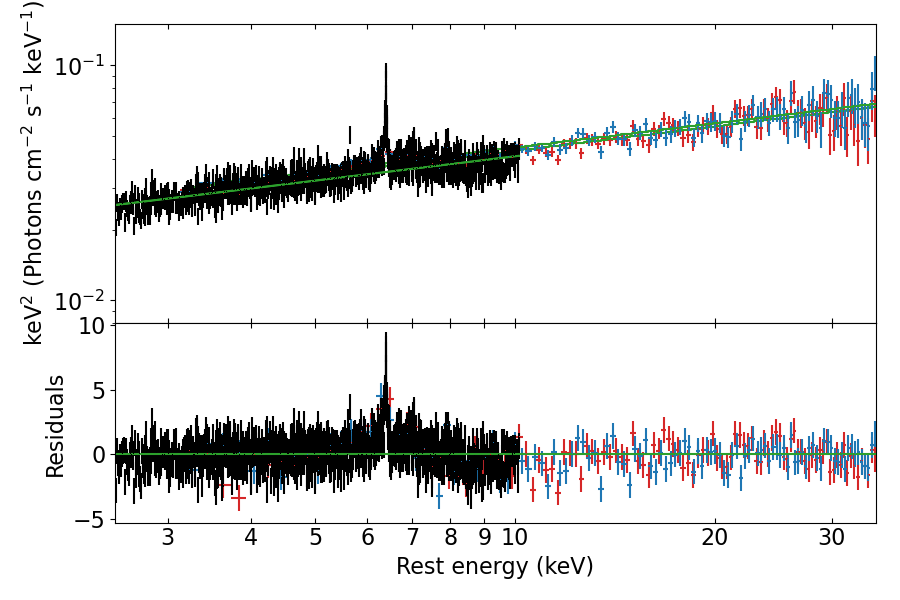}
    \includegraphics[width=\columnwidth]{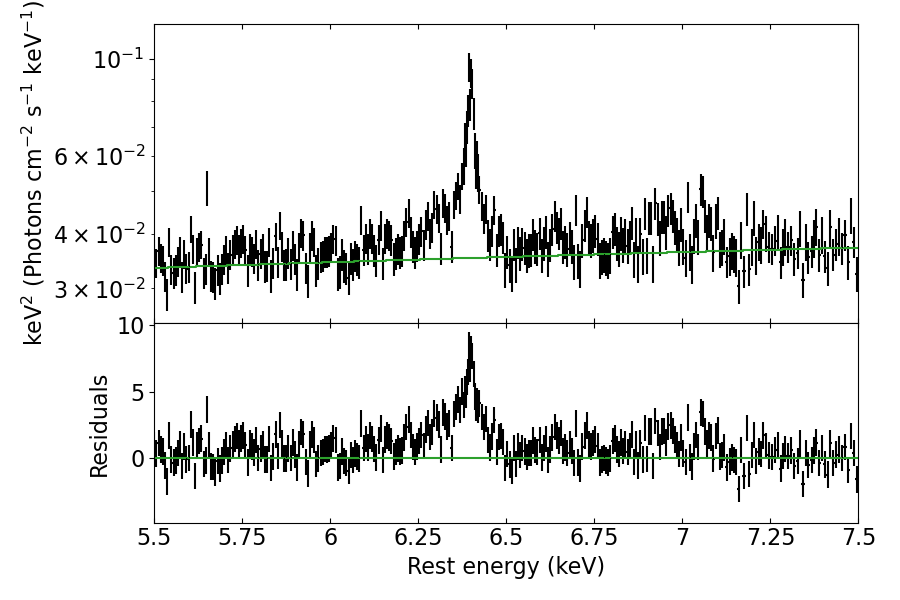}
            \caption{Broadband X-ray spectra of IC~4329A overlaid with the baseline power-law continuum model. Black, red, and blue points represent the \textit{XRISM}/Resolve, \textit{NuSTAR}/FPMA, and \textit{NuSTAR}/FPMB data, respectively. The lower panels show the residuals. The lower figure shows the inset around the Fe-K band.
            }
         \label{fig:continuum}
   \end{figure}

Figure~\ref{fig:continuum} displays the broad-band X-ray spectra overlaid with the baseline continuum model. The continuum is represented by a power law fitted outside the Fe-K band ($6$--$7~\mathrm{keV}$).
The best-fit photon index is $\Gamma \approx 1.7$, and the $2$--$10~\mathrm{keV}$ unabsorbed flux is $8.3 \times 10^{-11}~\mathrm{erg~cm^{-2}~s^{-1}}$, consistent with historical values for this source. 
%Following \citet{2024A&A...687A..51T}, who computed the bolometric luminosity from the $2\text{--}10~\mathrm{keV}$ flux using the luminosity-dependent bolometric correction of \citet{2010A&A...512A..34L} and adopted $M_{\mathrm{BH}} \approx 6.8 \times 10^7\ M_\odot$, our observed flux level corresponds to an Eddington ratio of $L / L_{\mathrm{Edd}} \approx 0.1$ (see their Figure~12).
Focusing on the $\mathrm{Fe\text{-}K}$ energy band, we clearly detect a narrow $\mathrm{Fe~K\alpha}$ core accompanied by a broader component that is slightly redshifted. Furthermore, the \ion{Fe}{25} forbidden line ($z$) at $6.64~\mathrm{keV}$, the \ion{Fe}{26}~$\mathrm{Ly}\alpha$ lines at $6.95$--$6.97~\mathrm{keV}$, and the $\mathrm{Fe~K\beta}$ emission line at $7.06~\mathrm{keV}$ are resolved.
Outside the $\mathrm{Fe\text{-}K}$ band, the spectrum is well described by a power-law continuum; notably, the Compton hump in the \textit{NuSTAR} band is remarkably weak.

We performed detailed spectral fitting within the \textit{XRISM}/Resolve bandpass ($2.4$--$12~\mathrm{keV}$). 
The narrow $\mathrm{Fe~K\alpha}$ and $\mathrm{Fe~K\beta}$ emission lines from the distant cold material are modeled using the \textsc{xclumpy} model convolved with a Gaussian velocity profile (\textsc{gsmooth}). 
The \textsc{xclumpy} model was originally developed by \citet{2019ApJ...877...95T} and subsequently extended to cover high-resolution Resolve spectra \citep{2026NatAs..10..880X}. The Fe K$\alpha$ line is modeled by seven Lorentzians \citep{holzer1997}.
We adopted the solar abundance table from \citet{lpgs}. 
The parameters are equatorial column density $N_{\mathrm{H,eq,torus}}$, torus angular width $\sigma_\mathrm{torus}$, inclination angle $\theta_\mathrm{i,torus}$, and photon index $\Gamma$. 
Here we assume that the torus and BLR share the same inclination angle (40~deg; \citealt{bentz2023}) and the torus angular width is 22.1~deg based on the X-ray broad band fitting \citep{2021ApJ...906...84O}.
Next, to account for the highly ionized \ion{Fe}{25} and \ion{Fe}{26} lines, positive Gaussian profiles taking into account cosmological redshift were included. 
For the \ion{Fe}{25} triplet, only the forbidden line ($z$) was incorporated, as adding the intercombination ($x+y$) and resonance ($w$) lines yielded normalizations consistent with zero within the $1\sigma$ confidence level. 
For \ion{Fe}{26}~$\mathrm{Ly}\alpha_{1,2}$, individual Gaussian components were introduced, with the normalization of $\mathrm{Ly}\alpha_2$ tied to half that of $\mathrm{Ly}\alpha_1$, and their line widths linked together. 
The line widths of \ion{Fe}{25} and \ion{Fe}{26} were fitted independently. 
Additionally, two ultra-fast outflow (UFO) zones manifest as absorption features at $\sim 8.5~\mathrm{keV}$ and $\sim 9.5~\mathrm{keV}$. 
{It should be noted that \citet{2006ApJ...646..783M} also reported a UFO absorption line but different energy, at 7.68 keV (rest frame), with $v = 0.093c$ and $\log \xi=3.7$, using the 2003 XMM-Newton observation.}
Because the continuum of IC~4329A is well approximated by a power law, we employed the photoionized plasma absorption table generated by \citet{mochizuki2023} using \texttt{XSTAR} v2.58e \citep{xstar}, assuming a power-law SED with $\Gamma = 1.8$. 
In this model, the column density ($N_{\mathrm{H}}$), ionization parameter ($\log\xi$), outflow velocity ($v_{\mathrm{wind}}$), and velocity dispersion ($\sigma_v$) were treated as free parameters. 
%Given their substantial blueshifts, these outflows are assumed to act only on the power-law continuum component. 
{While we can see a hint of absorption lines at $\sim7.0$~keV, the residuals do not show highly significant deviations, with $\Delta C = -14.9$ (see also \S\ref{sec:UFO}).}

The central objective of this work is the physical interpretation of the broad, redshifted $\mathrm{Fe~K\alpha}$ feature. We evaluated three candidate scenarios: (1) a redshifted xBLR, (2) relativistic accretion disk reflection, and (3) an unshifted xBLR combined with a mildly ionized absorber.

For all models considered below, the expected $C$-statistic is
$C_{\rm e} = 3113 \pm 79$, which is essentially determined by the number
of spectral bins and is insensitive to the model details.

\subsubsection{Redshifted xBLR model}
First, we modeled the broad $\mathrm{Fe~K\alpha}$ emission using an \textsc{xBLR} model convolved with a Gaussian kernel. 
The \textsc{xBLR} model was originally developed by \citet{2026ApJ..1003....6F} and Y.~Nakatani et al.\ submitted, and was tailored to match the parameters of IC~4329A. In this model,  a smooth, partially spherical geometry with two polar cuts is assumed. 
Based on optical reverberation mapping constraints \citep{bentz2023}, the BLR inclination angle $\theta_{i,\mathrm{BLR}}$ was fixed to $40^\circ$ and the half opening angle was fixed to $69^\circ$ (corresponding to a BLR angular width of $\sigma_\mathrm{BLR}=21^\circ$). 
To accommodate the redshifted broad feature, a velocity shift $v_{\mathrm{shift}}$ was introduced.
The equatorial column density $N_{\mathrm{H,eq,BLR}}$, velocity shift $v_{\mathrm{shift}}$, and velocity dispersion $\sigma_v$ were treated as free parameters for the xBLR model. 
The model function is 
\begin{equation}
\begin{split}
&\text{cross norm} * \\
&(\text{UFO1} * \text{UFO2} * \text{zcutoffpl}  + \\
&\text{gsmooth}*\text{xclumpy} + 
\text{gsmooth}*\text{xBLR} + \\
&\text{zgauss}_{\text{Ly}\alpha_\text{1}}+ 
\text{zgauss}_{\text{Ly}\alpha_\text{2}}+ 
\text{zgauss}_{\text{He}\alpha\text{,z}}.
\end{split}
\end{equation}

\begin{figure}[tb]
   \centering
    \includegraphics[width=\columnwidth]{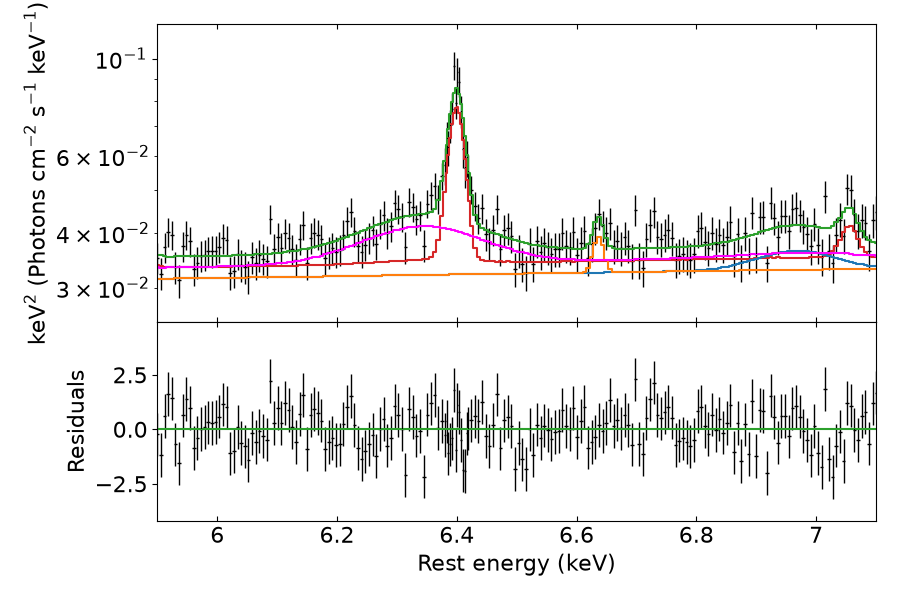}
    \includegraphics[width=\columnwidth]{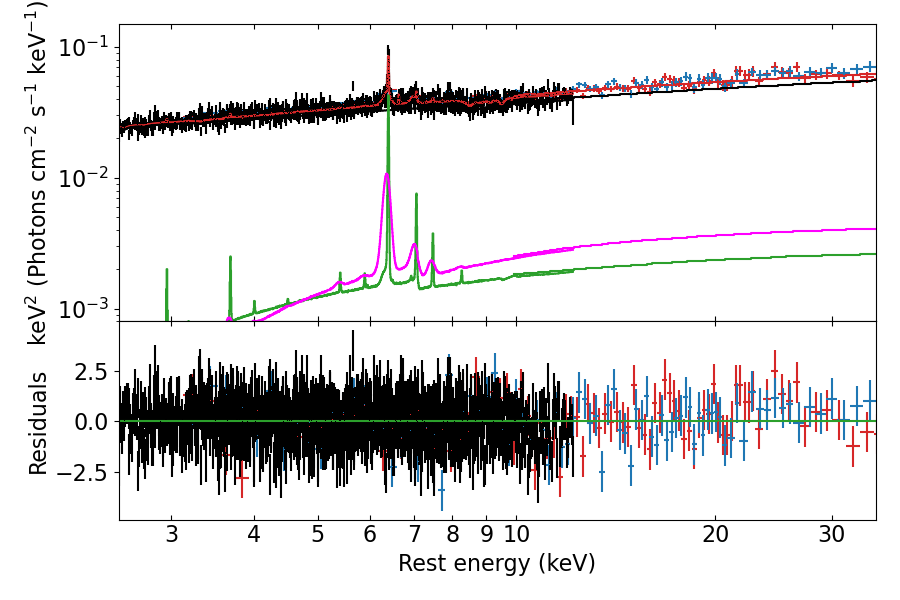}
    \includegraphics[width=\columnwidth]{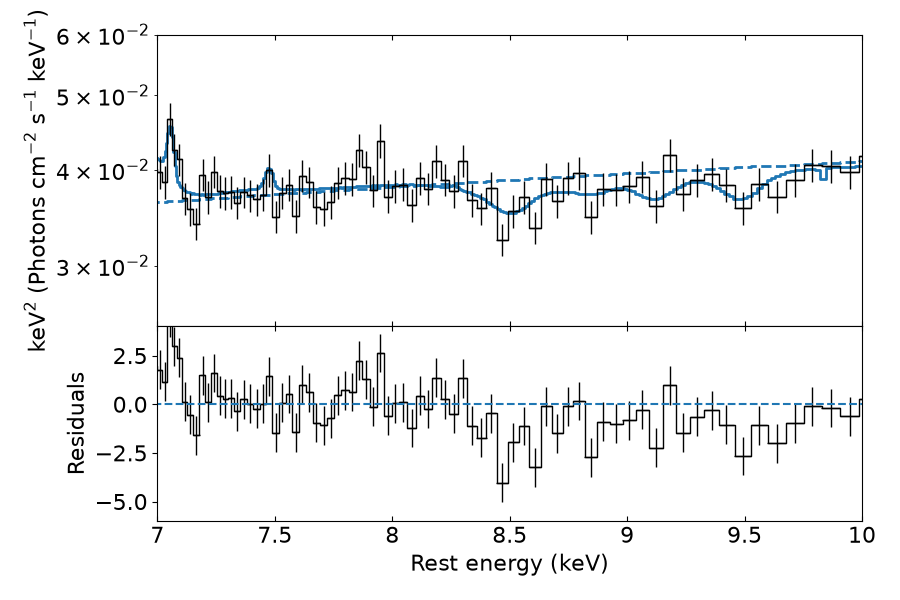}
            \caption{X-ray spectra of IC~4329A fitted with the redshifted xBLR model. 
(Top) \textit{XRISM}/Resolve spectrum in the $\mathrm{Fe~K}$ band. Green, magenta, and blue lines represent the torus reflection, redshifted xBLR emission, and ionized Fe emission lines, respectively, added to the power-law continuum. The red line indicates the total model. 
(Middle) Broadband fit extended to the \textit{NuSTAR} FPMA (red) and FPMB (blue) data. 
(Bottom) Zoomed-in view of the UFO absorption features in the 7.5--10~keV range. The blue solid line shows the best-fit model including UFO absorbers, while the blue dashed line illustrates the continuum model without UFO absorption. {The residuals against the continuum model is shown in the lower panel.}
            }
         \label{fig:redshiftedBLR}
   \end{figure}

The best-fit parameters are summarized in Table~\ref{tab:redshiftedBLR} and shown in Figure~\ref{fig:redshiftedBLR}. The width of the narrow $\mathrm{Fe~K\alpha}$ core corresponds to a velocity dispersion of $\sigma_v = 400_{-130}^{+180}~\mathrm{km~s^{-1}}$ ($\mathrm{FWHM} = 940_{-310}^{+420}~\mathrm{km~s^{-1}}$). 
The broad $\mathrm{Fe~K\alpha}$ component exhibits a significant redshift of $v_{\mathrm{shift}} = +1700_{-800}^{+1700}~\mathrm{km~s^{-1}}$, which is slightly larger than the one in the optical data.
Its  velocity dispersion is $\sigma_v = 3400_{-700}^{+1100}~\mathrm{km~s^{-1}}$ ($\mathrm{FWHM} = 8000_{-1600}^{+2600}~\mathrm{km~s^{-1}}$), which is consistent with or slightly smaller than the one in the optical data. 
The fit quality is good: the observed $C$-statistic of 3074.05 for 3089
degrees of freedom is fully consistent with the expected value of
$C_{\rm e} = 3113 \pm 79$, deviating by only $-0.5\sigma$. 

The lower panel in Figure~\ref{fig:redshiftedBLR} shows the UFO features.
To account for the broad negative residuals spanning 8.5--9.5~keV, it was necessary to introduce at least two distinct velocity components of relatively broad UFO absorbers.
Both components have ionization parameters of $\log(\xi / \mathrm{erg~cm~s^{-1}}) \sim 3$, where absorption features from \ion{Fe}{25} predominate.
%The upper limits for the velocity dispersions in both UFO and column density in UFO1 could not be constrained, as their upper uncertainty bounds hit the grid limit of $9000~\mathrm{km~s^{-1}}$ and $10^{23}$~cm$^{-2}$, respectively.
Compared to UFOs previously detected in local Seyfert 1 galaxies \citep{yamada2024}, these absorbers exhibit relatively low $\log\xi$ and modest $N_{\mathrm{H}}$.
Their statistical significance is discussed in Section~\ref{sec:UFO}.

\begin{table}[tb]
\centering
\caption{Best-fit spectral parameters for the redshifted xBLR model.}
\label{tab:redshiftedBLR}
\begin{tabular}{lll}
\hline\hline
Component & Parameter & Best-fit Value \\
\hline
Continuum & $\Gamma$ & $1.670_{-0.013}^{+0.012}$ \\
 & $\mathrm{norm}$ & $(1.88\pm0.03) \times 10^{-2}$ \\
\hline
Torus (\textsc{xclumpy}) & $N_{\mathrm{H,eq,torus}}\ (\mathrm{cm^{-2}})$ & $(2.2_{-0.5}^{+0.9}) \times 10^{23}$ \\
 & $\sigma_\mathrm{torus}\ (\mathrm{deg})$ & 22.1 (fixed) \\
 & $\theta_\mathrm{i,torus}\ (\mathrm{deg})$ & 40 (fixed) \\
 & $v_{\mathrm{shift}}\ (\mathrm{km~s^{-1}})$ & 0 (fixed) \\
 & $\sigma_v\ (\mathrm{km~s^{-1}})$ & $400_{-130}^{+180}$ \\
\hline
xBLR & $N_{\mathrm{H,eq,BLR}}\ (\mathrm{cm^{-2}})$ & $(3.3_{-1.4}^{+1.0}) \times 10^{23}$ \\
 & $\sigma_{\mathrm{BLR}}\ (\mathrm{deg})$ & 21 (fixed) \\
 & $\theta_\mathrm{i,BLR}\ (\mathrm{deg})$ & 40 (fixed) \\
 & $v_{\mathrm{shift}}\ (\mathrm{km~s^{-1}})$ & $+1700_{-800}^{+1700}$ \\
 & $\sigma_v\ (\mathrm{km~s^{-1}})$ & $3400_{-700}^{+1100}$ \\
\hline
\ion{Fe}{25} He$\alpha$,$z$ & $E_c\ (\mathrm{keV})$ & 6.63661 (fixed) \\
 & $\sigma_v\ (\mathrm{km~s^{-1}})$ & $400_{-200}^{+3900}$ \\
 & $v_{\mathrm{shift}}\ (\mathrm{km~s^{-1}})$ & 0 (fixed) \\
 & $\mathrm{norm}$ & $(3.4_{-2.0}^{+8.5}) \times 10^{-6}$ \\
\hline
\ion{Fe}{26}~$\mathrm{Ly}\alpha_1$ & $E_c\ (\mathrm{keV})$ & 6.97318 (fixed) \\
 & $\sigma_v\ (\mathrm{km~s^{-1}})$ & $2800_{-2000}^{+4000}$ \\
 & $v_{\mathrm{shift}}\ (\mathrm{km~s^{-1}})$ & 0 (fixed) \\
 & $\mathrm{norm}$ & $(8_{-5}^{+6}) \times 10^{-6}$ \\
\ion{Fe}{26}~$\mathrm{Ly}\alpha_2$ & $E_c\ (\mathrm{keV})$ & 6.95197 (fixed) \\
 & $\sigma_v\ (\mathrm{km~s^{-1}})$ & Same as $\mathrm{Ly}\alpha_1$ \\
 & $v_{\mathrm{shift}}\ (\mathrm{km~s^{-1}})$ & 0 (fixed) \\
 & $\mathrm{norm}$ & $0.5 \times \mathrm{norm}(\mathrm{Ly}\alpha_1)$ \\
\hline
UFO 1 & $N_{\mathrm{H}}\ (\mathrm{cm^{-2}})$ & {$(6_{-4}^{+15}) \times 10^{22}$} \\
 & $\log(\xi / \mathrm{erg~cm~s^{-1}})$ & $3.16_{-0.20}^{+0.18}$ \\
 & $\sigma_v\ (\mathrm{km~s^{-1}})$ & {$6_{-5}^{+4}\times10^3$} \\
 & $v_{\mathrm{wind}}/c$ & $-0.300 \pm 0.005$ \\
\hline
UFO 2 & $N_{\mathrm{H}}\ (\mathrm{cm^{-2}})$ & $(3.3_{-1.7}^{+2.9}) \times 10^{22}$ \\
 & $\log(\xi / \mathrm{erg~cm~s^{-1}})$ & $2.95_{-0.17}^{+0.11}$ \\
 & $\sigma_v\ (\mathrm{km~s^{-1}})$ & {$(8\pm3)\times10^3$} \\
 & $v_{\mathrm{wind}}/c$ & $-0.235_{-0.003}^{+0.005}$ \\
\hline
Const. & $C_{\mathrm{FPMA}} / C_{\mathrm{Resolve}}$ & $1.039\pm0.010$ \\
 & $C_{\mathrm{FPMB}} / C_{\mathrm{Resolve}}$ & $1.073\pm0.010$ \\
\hline
\multicolumn{2}{l}{$C$-statistic / d.o.f.} & 3074.05 / 3089 \\
\multicolumn{2}{l}{Expected $C$-statistic} & $3113 \pm 79$ \\
\hline
\end{tabular}
\end{table}

\subsubsection{Disk line model}
Next, we replaced the redshifted xBLR component with a relativistic disk reflection model, \textsc{relxill} v2.9.0 \citep{2014MNRAS.444L.100D, 2014ApJ...782...76G}.
The emissivity index, black hole spin, outer disk radius, and iron abundance were fixed to $q = -3$, $a = 0$, $R_{\mathrm{out}} = 400\,R_g$, and $A_{\mathrm{Fe}} = 1$, respectively.
The disk inclination angle $\theta_{i,\mathrm{disk}}$, inner disk radius $R_{\mathrm{in}}$, ionization parameter $\log\xi$, and reflection fraction $R_{\mathrm{refl}}$ were allowed to vary.
The parameters for the highly ionized emission lines and UFO absorbers were fixed to the values obtained in the redshifted xBLR model.
The model function is 

\begin{equation}
\begin{split}
&\text{cross norm} * \\
&(\text{UFO1} * \text{UFO2} * (\text{zcutoffpl} + \text{relxill}) + \\
&\text{gsmooth}*\text{xclumpy} + \\
&\text{zgauss}_{\text{Ly}\alpha_\text{1}}+ 
\text{zgauss}_{\text{Ly}\alpha_\text{2}}+ 
\text{zgauss}_{\text{He}\alpha\text{,z}}.
\end{split}
\end{equation}

\begin{figure}[tb]
   \centering
    \includegraphics[width=\columnwidth]{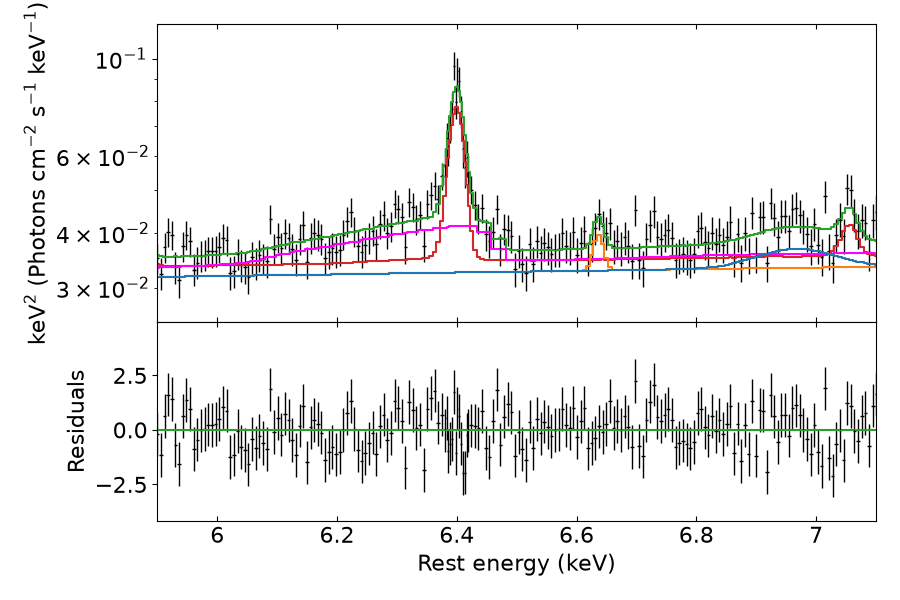}
            \caption{Same as Figure~\ref{fig:redshiftedBLR} (top), but for the relativistic disk reflection model (\textsc{relxill}). The orange line represents the disk line reflection profile.
            }
         \label{fig:diskline}
   \end{figure}

The inner disk radius $R_{\mathrm{in}}$ is primarily constrained by the extent of the red wing, whereas the blue wing cutoff ($\sim 6.5~\mathrm{keV}$) is jointly determined by $R_{\mathrm{in}}$ and $\theta_{i,\mathrm{disk}}$. The fit favors a low inclination angle of $\theta_{i,\mathrm{disk}} \approx 3^\circ$ (Table~\ref{tab:diskline} and Figure~\ref{fig:diskline}). This value differs substantially from the BLR inclination angle of $40^\circ$ derived from optical reverberation mapping, implying a warped disk structure if this scenario holds. Furthermore, because $R_{\mathrm{in}} \approx 50\,R_g$ does not extend down to the innermost stable circular orbit, the inner accretion disk may either be truncated or fully ionized, suppressing iron fluorescence in the innermost region \citep[e.g.,][]{2026ApJ..1005..119N}.
The fit quality is also good ($C = 3076.42$ for 3088 d.o.f., compared with
$C_{\rm e} = 3113 \pm 79$; $-0.5\sigma$).
We note that fixing the spin parameter at $a=0.998$ (maximum spin Kerr BH) does not significantly change the results, including $R_{\mathrm{in}}$.

\begin{table}[tb]
\centering
\caption{Best-fit spectral parameters for the relativistic disk reflection model.}
\label{tab:diskline}
\begin{tabular}{lll}
\hline\hline
Component & Parameter & Best-fit Value \\
\hline
Continuum & $\Gamma$ & $1.69_{-0.01}^{+0.01}$ \\
 & $\mathrm{norm}$ & $(1.98 \pm 0.03) \times 10^{-2}$ \\
\hline
Torus (\textsc{xclumpy}) & $N_{\mathrm{H,eq,torus}}\ (\mathrm{cm^{-2}})$ & $(3.7_{-0.5}^{+0.6})\times10^{23}$ \\
 & $\sigma_\mathrm{torus}\ (\mathrm{deg})$ & 22.1 (fixed) \\
 & $\theta_\mathrm{i,torus}\ (\mathrm{deg})$ & 40 (fixed) \\
 & $v_{\mathrm{shift}}\ (\mathrm{km~s^{-1}})$ & 0 (fixed) \\
 & $\sigma_v\ (\mathrm{km~s^{-1}})$ & $690\pm120$ \\
\hline
Disk Reflection & $\theta_{i,\mathrm{disk}}\ (\mathrm{deg})$ & $3.0_{-3.0}^{+3.4}$ \\
(\textsc{relxill}) & $R_{\mathrm{in}}\ (R_g)$ & $54_{-12}^{+11}$ \\
 & $\log(\xi / \mathrm{erg~cm~s^{-1}})$ & $0.0_{-0.0}^{+1.9}$ \\
 & $A_\mathrm{Fe}$ & 1 (fixed) \\
 & reflection fraction & $0.138\pm0.018$ \\
\hline
Const. & $C_{\mathrm{FPMA}} / C_{\mathrm{Resolve}}$ & $1.036\pm0.009$ \\
 & $C_{\mathrm{FPMB}} / C_{\mathrm{Resolve}}$ & $1.069\pm0.010$ \\
 \hline
\multicolumn{2}{l}{$C$-statistic / d.o.f.} & 3076.42 / 3088 \\
\multicolumn{2}{l}{Expected $C$-statistic} & $3113 \pm 79$ \\
\hline
\end{tabular}
%\begin{tablenotes}
%\item $\dagger$ Ionized emission lines and UFO parameters were fixed to the values in Table~\ref{tab:redshiftedBLR}.
%\end{tablenotes}
\end{table}

\subsubsection{Absorbed xBLR model}
Third, we considered a scenario where the blue wing of an unshifted broad $\mathrm{Fe~K\alpha}$ line is partially suppressed by a mildly ionized absorber. A similar feature was detected in NGC~4388 by \citet{2026ApJ..1003....6F}, where an absorption feature near $6.5~\mathrm{keV}$ distorted the xBLR profile. Following this result, we applied a mildly ionized absorber to an unshifted xBLR component.

\begin{figure}[tb]
   \centering
    \includegraphics[width=\columnwidth]{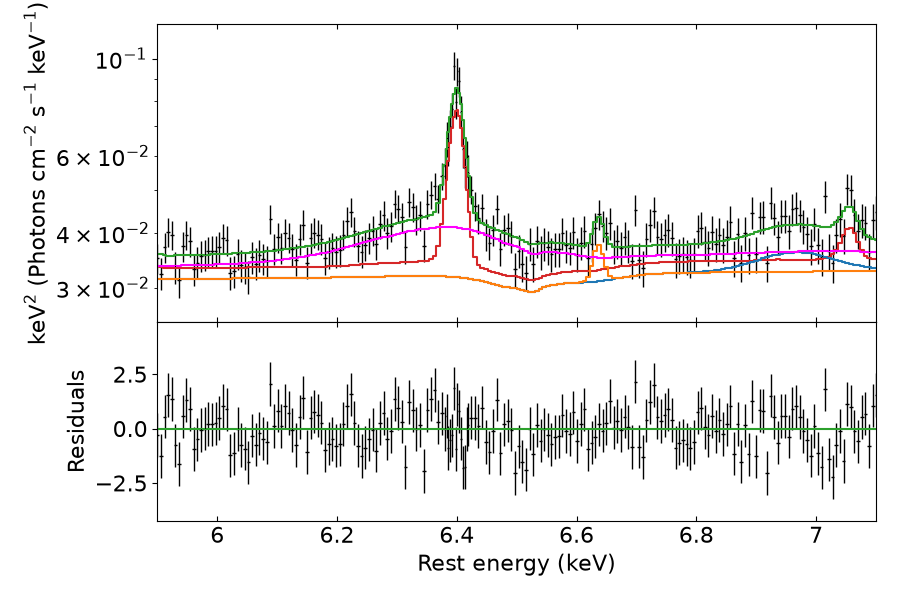}
            \caption{Same as Figure \ref{fig:redshiftedBLR} (top), but for the absorbed BLR model. 
            }
         \label{fig:BLR_abs}
   \end{figure}

This fitting also yields a statistically good fit ($C = 3083.35$ for
3086 d.o.f., compared with $C_{\rm e} = 3113 \pm 79$; $-0.4\sigma$;
Table~\ref{tab:BLR_abs} and Figure~\ref{fig:BLR_abs}).
The velocity of the mildly ionized absorber is $+900_{-800}^{+600}$~km~s$^{-1}$; it has no radial motion or is slightly inflowing.
The column density and ionization parameter are consistent with those of typical warm absorbers \citep[e.g.,][]{yamada2024}.

\begin{table}[tb]
\centering
\caption{Best-fit spectral parameters for the absorbed xBLR model.}
\label{tab:BLR_abs}
\begin{tabular}{lll}
\hline\hline
Component & Parameter & Best-fit Value \\
\hline
Continuum & $\Gamma$ & $1.68_{-0.01}^{+0.02}$ \\
 & $\mathrm{norm}$ & $(1.92_{-0.04}^{+0.05}) \times 10^{-2}$ \\
\hline
Torus (\textsc{xclumpy}) & $N_{\mathrm{H,eq,torus}}\ (\mathrm{cm^{-2}})$ & $(1.9_{-0.5}^{+0.7})\times10^{23}$ \\
 & $\sigma_\mathrm{torus}\ (\mathrm{deg})$ & 22.1 (fixed) \\
 & $\theta_\mathrm{i,torus}\ (\mathrm{deg})$ & 40 (fixed) \\
 & $v_{\mathrm{shift}}\ (\mathrm{km~s^{-1}})$ & 0 (fixed) \\
 & $\sigma_v\ (\mathrm{km~s^{-1}})$ & $340_{-130}^{+100}$ \\
\hline
xBLR & $N_{\mathrm{H,eq,BLR}}\ (\mathrm{cm^{-2}})$ & $(3.8_{-1.0}^{+1.2}) \times 10^{23}$ \\
 & $\sigma_{\mathrm{BLR}}\ (\mathrm{deg})$ & 21 (fixed) \\
 & $\theta_\mathrm{i,BLR}\ (\mathrm{deg})$ & 40 (fixed) \\
 & $v_{\mathrm{shift}}\ (\mathrm{km~s^{-1}})$ & 0 (fixed) \\
 & $\sigma_v\ (\mathrm{km~s^{-1}})$ & $3300_{-1000}^{+2300}$ \\
\hline
Mildly Ionized & $N_{\mathrm{H}}\ (\mathrm{cm^{-2}})$ & $(0.6_{-0.3}^{+0.5}) \times 10^{22}$ \\
Absorber & $\log(\xi / \mathrm{erg~cm~s^{-1}})$ & $2.45_{-0.20}^{+0.06}$ \\
 & $\sigma_v\ (\mathrm{km~s^{-1}})$ & $1000_{-600}^{+2200}$ \\
 & $v_{\mathrm{wind}}\ (\mathrm{km~s^{-1}})$ & $+900_{-800}^{+600}$ \\
\hline
Const. & $C_{\mathrm{FPMA}} / C_{\mathrm{Resolve}}$ & $1.039\pm0.010$ \\
 & $C_{\mathrm{FPMB}} / C_{\mathrm{Resolve}}$ & $1.073\pm0.010$ \\
\hline
\multicolumn{2}{l}{$C$-statistic / d.o.f.} & 3083.35 / 3086 \\
\multicolumn{2}{l}{Expected $C$-statistic} & $3113 \pm 79$ \\
\hline
\end{tabular}
\end{table}

\subsubsection{Hybrid model, or optically constrained xBLR and disk line model} \label{sec:hybrid}
Finally, we assumed that the optical/NIR BLR and xBLR share identical velocity widths and velocity shifts.
Based on Section~\ref{sec:result_optical}, the FWHM was fixed at $6000~\mathrm{km~s^{-1}}$, while the velocity shift ($v_{\mathrm{shift}}$) was allowed to vary within the range of $600\text{--}1000~\mathrm{km~s^{-1}}$ during the spectral fitting.
Under this assumption, the Fe-K line profile cannot be fully reproduced; thus, we also introduced the relativistic disk reflection component.
The model function corresponds to Equation~(1) with the addition of the \textsc{relxill} component.

\begin{figure}[tb]
   \centering
    \includegraphics[width=\columnwidth]{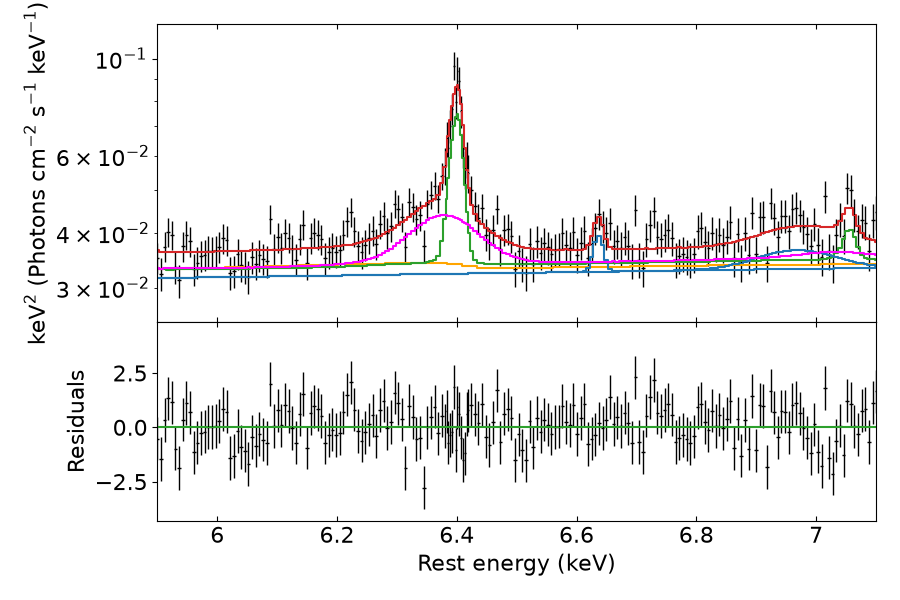}
            \caption{Same as Figure \ref{fig:redshiftedBLR} (top), but for the hybrid model. The magenta and orange lines represent the xBLR and disk reflection components, respectively.
            }
         \label{fig:hybrid}
   \end{figure}

Table~\ref{tab:hybrid} and Figure~\ref{fig:hybrid} show the fitting results.
The EW of the broad component is predominantly accounted for by the xBLR component, with the remaining residuals being filled by a strongly redshift-skewed disk line component.
To accommodate the required stronger gravitational redshift, the inner radius ($R_{\mathrm{in}}$) shifts to a smaller value compared to that in the disk-line model.
The inclination angle remains consistent with a nearly face-on geometry.
The torus component has $\text{FWHM}=870_{-280}^{+260}$~km~s$^{-1}$.

The hybrid model yields $C = 3044.90$ for 3086 d.o.f., which is again
consistent with $C_{\rm e} = 3113 \pm 79$ ($-0.9\sigma$).
Although the $C$-statistic improves by $\sim 30$--$40$ relative to the other
models, the observed $C$-statistics of all four models lie within $1\sigma$
of the expected value (Tables~\ref{tab:redshiftedBLR}--\ref{tab:hybrid}).
Consequently, distinguishing between these models is not feasible based
solely on the current X-ray spectral fitting.
We note that these models are not nested, so the $\Delta C$ values cannot be
directly converted into formal significances.

\begin{table}[htbp]
\centering
\caption{Best-fit spectral parameters for the hybrid model.}
\label{tab:hybrid}
\begin{tabular}{lll}
\hline\hline
Component & Parameter & Best-fit Value \\
\hline
Continuum & $\Gamma$ & $1.70\pm0.01$ \\
 & $\mathrm{norm}$ & $(1.94\pm0.04) \times 10^{-2}$ \\
\hline
Torus (\textsc{xclumpy}) & $N_{\mathrm{H,eq,torus}}\ (\mathrm{cm^{-2}})$ & $(2.1_{-0.3}^{+0.5})\times10^{23}$ \\
 & $\sigma_\mathrm{torus}\ (\mathrm{deg})$ & 22.1 (fixed) \\
 & $\theta_\mathrm{i,torus}\ (\mathrm{deg})$ & 40 (fixed) \\
 & $v_{\mathrm{shift}}\ (\mathrm{km~s^{-1}})$ & 0 (fixed) \\
 & $\sigma_v\ (\mathrm{km~s^{-1}})$ & $370_{-120}^{+110}$ \\
\hline
xBLR & $N_{\mathrm{H,eq,BLR}}\ (\mathrm{cm^{-2}})$ & $(2.8_{-0.8}^{+1.1}) \times 10^{23}$ \\
 & $\sigma_{\mathrm{BLR}}\ (\mathrm{deg})$ & 21 (fixed) \\
 & $\theta_\mathrm{i,BLR}\ (\mathrm{deg})$ & 40 (fixed) \\
 & $v_{\mathrm{shift}}\ (\mathrm{km~s^{-1}})$ & $+940$ \\
 & $\sigma_v\ (\mathrm{km~s^{-1}})$ & $2550$ (fixed) \\
\hline
Disk Reflection & $\theta_{i,\mathrm{disk}}\ (\mathrm{deg})$ & $3_{-3}^{+13}$ \\
(\textsc{relxill}) & $R_{\mathrm{in}}\ (R_g)$ & $13.2_{-2.3}^{+5.5}$ \\
 & $\log(\xi / \mathrm{erg~cm~s^{-1}})$ & $1.7_{-1.7}^{+0.4}$ \\
 & $A_\mathrm{Fe}$ & 1 (fixed) \\
 & reflection fraction & $0.13\pm0.02$ \\
\hline
Const. & $C_{\mathrm{FPMA}} / C_{\mathrm{Resolve}}$ & $1.037\pm0.009$ \\
 & $C_{\mathrm{FPMB}} / C_{\mathrm{Resolve}}$ & $1.071_{-0.009}^{+0.010}$ \\
\hline
\multicolumn{2}{l}{$C\text{-statistic} / \text{d.o.f.}$} & 3044.90 / 3086 \\
\multicolumn{2}{l}{Expected $C$-statistic} & $3113 \pm 79$ \\
\hline
\end{tabular}
\end{table}

\section{Discussion} \label{sec:discussion}
\subsection{Plausible origin of the red wing in the Fe K$\alpha$ line}
We evaluated four candidate scenarios: (1) a redshifted xBLR,
(2) relativistic accretion disk reflection,  (3) an unshifted xBLR
combined with a mildly ionized absorber, and (4) a hybrid model.
While all the models provide statistically acceptable fits to the X-ray spectrum alone, the redshifted xBLR scenario or the hybrid model offer the most physically self-consistent picture when integrated with our optical and NIR spectroscopic findings.
The primary support for the redshifted xBLR origin is the consistent redshift detected across multiple wavelength regimes. Both optical ($\mathrm{H}\beta$) and NIR ($\mathrm{Pa}\beta$) broad emission lines show velocity redshifts of $v \approx +600$--$1000~\mathrm{km~s^{-1}}$ (Section~\ref{sec:result_optical}). It is natural to infer that the xBLR gas originates from the same dynamical structure responsible for the optical/NIR broad lines. Recent comparative analyses combining \textit{XRISM} data with ground-based optical/NIR spectroscopy have similarly revealed matching velocity profiles between xBLR and optical/NIR BLRs in other AGN \citep[e.g.,][H.~Noda et al.\ in prep.]{2026ApJ...997L..26Z}. While the other scenarios cannot be ruled out, one would be forced to invoke an ad hoc explanation for why the centroid velocity of the broad $\mathrm{Fe~K\alpha}$ line differs markedly from its optical/NIR counterparts.

Asymmetric and shifted broad-line profiles are increasingly recognized as a ubiquitous feature in AGNs, particularly in hard X-ray-selected sources operating at moderate-to-low accretion rates \citep{2025ApJ...991..116W}. Such asymmetries are commonly attributed to Keplerian motion within non-axisymmetric accretion structures, such as spiral density waves, localized hot spots, or asymmetric wind outflows. Furthermore, optical and NIR interferometric studies by \citet{2024A&A...684A.167G} have resolved distinct spatial photocenter shifts and velocity offsets within the BLR in IC 4329A, pointing toward strong dynamical influences from radiation pressure and non-axisymmetric gas distributions.

\subsection{Size scale of xBLR}

To quantitatively evaluate the scale of the xBLR, we estimate its characteristic radius ($R_{\mathrm{BLR}}$) from the broad $\mathrm{Fe~K\alpha}$ line width. The physical radius is expressed as:
\begin{eqnarray}
    R_\mathrm{BLR}&=&GM_\mathrm{BH}/v_\mathrm{rot}^2 \\
    v_\mathrm{rot} &=& f\times(\text{FWHM}) \label{eq:fudge}
\end{eqnarray}
where 
$f$ is a fudge factor that depends on the assumed BLR model \citep{2008MNRAS.386L..15D}. 
%For example, when BLR is 2D Keplerian disk, the observed FWHM is
%$2 v_\mathrm{rot} \sin \theta_\mathrm{i,BLR}$ so that $f=(2\sin\theta_\mathrm{i,BLR})^{-1}$ .
Here, 
denoting $v_{\rm turb}$ as the one-dimensional (line-of-sight) turbulent velocity dispersion, the net observable velocity width can be written as
\begin{equation}
v_{\mathrm{obs}}^2 = v_{\mathrm{rot}}^2 \sin^2 \theta_\mathrm{i,BLR} + v_{\mathrm{turb}}^2 \label{eq:turb}
\end{equation} 
In a disk with finite vertical scale height $H$, hydrostatic equilibrium relates the vertical (one-dimensional) turbulent velocity to the rotational velocity via $v_{\rm turb} \approx (H/R) v_{\rm rot}$ \citep{2008MNRAS.386L..15D}.
This yields
\begin{equation}
v_{\mathrm{obs}} = v_{\mathrm{rot}} \sqrt{ \sin^2 \theta_\mathrm{i,BLR} + (H/R)^2 }
\end{equation}
Identifying the observed velocity with $v_{\rm obs} = \mathrm{FWHM}/2$,
the fudge factor is calculated to be 
\begin{equation}
    f=\left[2\sqrt{\sin^2 \theta_\mathrm{i,BLR}+(H/R)^2}\right]^{-1},
\end{equation}
in agreement with Equation~(4) of \citet{2008MNRAS.386L..15D}.
With $\theta_\mathrm{i,BLR} = 40$~deg and $H/R=\tan(21^\circ)$, $f=0.67$.
Following the redshifted xBLR model, the line width is $\mathrm{FWHM} = 8000_{-1600}^{+2600}~\mathrm{km~s^{-1}}$. With $M_\mathrm{BH}=6.8 \times 10^7 M_{\odot}$,
$R_\mathrm{BLR}=(3.2_{-1.4}^{+1.8}) \times 10^3 \ R_g = 12_{-5}^{+7}\text{ light-days} = (1.0_{-0.4}^{+0.6}) \times 10^{-2}\text{ pc}$.
Based on the hybrid model, in which the line width is fixed at
$\mathrm{FWHM}=6000$~km~s$^{-1}$ (Section~\ref{sec:hybrid}),
$R_\mathrm{BLR}=5.6 \times 10^3\ R_g = 22\text{ light-days} = 1.8 \times 10^{-2}\text{ pc}$.

\subsection{Narrow Fe K$\alpha$ line} \label{sec:narrow}

The line width of the Fe K$\alpha$ emission was constrained using the Chandra/HETG data (see e.g., \citealt{2004ApJ...604...63Y,2004ApJ...608..157M}). We have revisited the HETG spectrum of IC 4329A, which combines six observations conducted between 2002 and 2018, yielding a total net exposure time of 230~ks. The spectrum in the 2--7.5~keV energy range was analyzed using a baseline model consisting of a power-law continuum and \texttt{zbfeklor}, which is a seven Lorentzian approximation model to the line profile of Fe K$\alpha$ from neutral material \citep{holzer1997} with line broadening and redshift. The best-fit results yielded an Fe K$\alpha$ line width of $\mathrm{FWHM} = 7700_{-2100}^{+2600}~\mathrm{km~s^{-1}}$. This value lies intermediate between the narrow and broad components resolved by XRISM. This indicates that due to the spectral resolution limits of the HETG, the two components were indistinguishable, resulting in a single line fit over an unresolved, blended profile.

Next, we discuss the radius of the torus. 
Assuming Equation~(\ref{eq:turb}) holds for the torus region as well, X-ray spectroscopy can probe not only the dense torus body itself but also sub-structures such as an inner fountain flow or a dusty wind (polar dust). According to the radiation fountain flow model, the outflow velocity reaches up to $500~\mathrm{km~s^{-1}}$ at smaller radii and decelerates to approximately $100~\mathrm{km~s^{-1}}$ toward the outer region \citep{2016ApJ...828L..19W}. Determining the observable line width of the Fe K$\alpha$ emission in a dynamical torus model requires detailed radiative transfer calculations. For simplicity, here we consider isotropic motion in the torus and adopt $H/R = \tan 22.1^\circ$ in Equation~(\ref{eq:turb}). Based on the hybrid model with $\text{FWHM}=870_{-280}^{+260}$~km~s$^{-1}$, the torus radius is estimated to be $R_\mathrm{torus}=(2.7_{-1.1}^{+3.2}) \times 10^5\ R_g = (1.1_{-0.4}^{+1.3}) \times 10^3\text{ light-days} = 0.9_{-0.4}^{+1.0}\text{ pc}$.

The inner radius of the dust torus ($R_{\rm dust}$) is determined by the dust sublimation temperature, $T_{\rm sub}\simeq1500$~K \citep{1987ApJ...320..537B}. Dust
evaporates when the irradiation flux reaches $L/(4\pi R^2) = \sigma_{\rm SB}T_{\rm sub}^4$,
where $\sigma_{\rm SB}$ is the Stefan-Boltzmann constant.
Therefore,
$R_{\rm dust}=(L/(4\pi\sigma_{\rm SB} T_{\rm sub}^4))^{1/2}$.
It is calculated to be $\simeq500$~light-days or 0.4~pc in IC 4329A \citep{2022A&A...664A..46A}.
When we simply assume a circular cross-section torus, it has a
self-shielding radius ($R_{\rm shield}$), as $R_{\rm shield}/R_{\rm
  dust}=\cos\theta/(1-\sin\theta)$, where $\theta$ is the torus angular width \citep{Mizumoto2019}.  In IC 4329A, $R_{\rm shield}=0.6$~pc.
Our estimate of $R_\mathrm{torus}$ is comparable to or slightly larger than
$R_{\rm shield}$ (consistent within the uncertainties), which provides a
consistent picture if the narrow Fe K$\alpha$ line broadly traces the torus structure.

\subsection{Significance of the UFO features} \label{sec:UFO}
\begin{figure}[tb]
   \centering
    \includegraphics[width=\columnwidth]{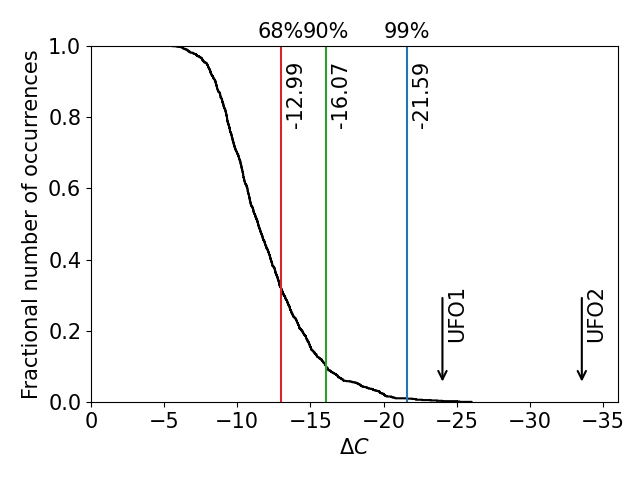}
            \caption{
            Empirical cumulative probability distribution of the improvement in fit statistic ($\Delta C$) obtained from 1000 simulated \textit{XRISM}/Resolve spectra under the null hypothesis (no UFO absorption). Vertical dashed lines indicate significance thresholds corresponding to different local null probabilities.
            }
         \label{fig:cumu}
   \end{figure}

To assess the statistical significance (or local null-hypothesis probability) of the candidate ultra-fast outflows (UFOs) while properly accounting for trial factors, we implemented a Look-Elsewhere Effect (LEE) correction framework \citep[e.g.,][]{kosec17}.
The trial factor was empirically calibrated by generating a synthetic ensemble of 1000 \textit{Resolve} spectra under the baseline continuum model devoid of any UFO absorption features. Subsequently, each simulated dataset was fitted with an additional \textsc{xstar} grid component, treating the column density, ionization parameter, and line velocity dispersion as free variables. We performed a systematic redshift scan across the interval $z \in [-0.334, +0.016]$ in 350 uniform steps, tracking the corresponding improvement in the fit statistic ($\Delta C$).
Figure~\ref{fig:cumu} displays the resulting empirical cumulative probability distribution. Based on this benchmark, a feature is considered detected at the 99\% significance level if the fit improvement satisfies $\Delta C < -21.59$. The inclusion of UFO1 and UFO2 yields $\Delta C = -24.03$ and $\Delta C = -33.56$, respectively. Thus, both features comfortably exceed the threshold, establishing their detection at a confidence level exceeding 99\% under the LEE criteria.

\section{Conclusion} \label{sec:conclusion}
\begin{figure*}[tb]
   \centering
    \includegraphics[width=1.8\columnwidth]{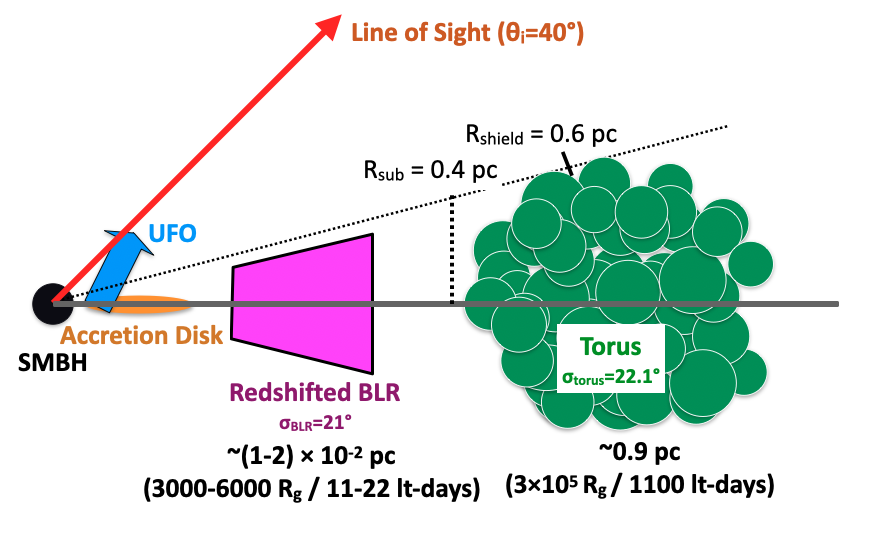}
            \caption{Schematic picture of the circumnuclear environment of IC 4329A, based on  the hybrid model.
            }
         \label{fig:schematic}
   \end{figure*}

We have performed multi-wavelength spectral analyses of the X-ray bright Seyfert 1 galaxy IC 4329A using XRISM/Resolve, NuSTAR, Seimei/KOOLS-IFU, and Magellan/WINERED data. Figure \ref{fig:schematic} shows the schematic picture of IC 4329A.

\begin{enumerate}
    \item Based  on the  X-ray spectrum from XRISM/Resolve and NuSTAR, four  physical scenarios were tested: (1) a redshifted xBLR model, (2) a relativistic disk reflection model, (3) an unshifted xBLR model affected by a mildly ionized absorber, and (4) a hybrid model. All provide statistically acceptable and comparable fits.
    
    \item As a result of quasi-simultaneous optical ($\mathrm{H}\beta$) and NIR ($\mathrm{Pa}\beta$) spectroscopy, the optical/NIR broad line shows redshifts of $v_{\rm shift} = +600-1000\text{ km s}^{-1}$. Thus, it is reasonable to expect that the xBLR exhibits a similar redshift.
    
\item The physical radius derived for the xBLR is $R_{\rm BLR} \sim (3.2\text{--}5.6) \times 10^3\text{ }R_g \sim 12\text{--}22\text{ light-days}$ (${\rm FWHM} \sim 6000\text{--}8000\text{ km s}^{-1}$). Meanwhile, the line width of the narrow Fe K$\alpha$ core in the hybrid model (${\rm FWHM} = 870_{-280}^{+260}\text{ km s}^{-1}$) places the torus reflection radius at $R_{\rm torus} \sim 1100\text{ light-days} \sim 0.9\text{ pc}$, tracing fluorescence across the torus structure beyond the dust sublimation ($R_{\rm sub} \sim 0.4\text{ pc}$) and self-shielding ($R_{\rm shield} \sim 0.6\text{ pc}$) boundaries. 
    
    \item Two robust UFO absorption components are resolved at $\sim 8.5\text{ keV}$ and $\sim 9.5\text{ keV}$, corresponding to ultra-fast outflow velocities of $v_{\rm wind}/c = -0.300 \pm 0.005$ and $-0.235_{-0.003}^{+0.005}$. Both absorption features exceed a 99\% confidence level threshold after accounting for trial factors via look-elsewhere effect Monte Carlo simulations.
\end{enumerate}

%% Please use the acknowledgment and contribution environments. This will 
%% be anonomyized when the "anonymous" style option is used. 
\begin{acknowledgments}
This research has made use of the \textit{NuSTAR} Data Analysis Software (NuSTARDAS) jointly developed by the ASI Space Science Data Center (SSDC, Italy) and the California Institute of Technology (Caltech, USA).
WINERED was developed by the University of Tokyo and the Laboratory of Infrared High-resolution spectroscopy (LiH), Kyoto Sangyo University.
The authors thank the staff of Las Campanas Observatory and Okayama Observatory for operations of the Seimei telescope and the Magellan telescope, respectively.
The authors also thank Scarlet Elgueta and Matilde Coello-Guzm\'{a}n for WINERED operation and the data collection at the Magellan telescope.

Part of this work was supported by 
JSPS KAKENHI Grant Number JP26K17197 (MM), JP26K17187 (SY), JP23K13153, JP24K00673 (TK) and
Yamada Science Foundation (MM).
Y.N. acknowledges financial support from Grant-in-Aid for JSPS Fellows (24KJ1507).

Machine learning tools were used to suggest language improvements within the manuscript.
Responsibility for the final manuscript text lies entirely with the authors.

The optical/NIR data used in this paper are archived in  \url{https://github.com/mskmizu/IC4329Apaper_data}. 
The X-ray data are archived in  the HEASARC web page.

\end{acknowledgments}

\begin{contribution}
%%This section gives authors the space to recognize author contributions. The text inside this environment is NOT counted towards the total word quanta. At a minimum, manuscripts are expected to include this text:

MM led the project and was responsible for writing and submitting the manuscript.
MM, YN, SY, YU, TK, SO, YM, and KF discussed interpretation of the \textit{XRISM} data.
MM, YN, SY, and FI performed the Seimei/KOOLS-IFU observations and its data reduction. SO, NM, DT, and HK performed the Magellan/WINERED observations and its data reduction.
All authors were responsible for the manuscript proofing.

%% But authors are expected to provide more specific details, e.g. 
%%
%%SC was responsible for writing and submitting the manuscript.
%%WWM came up with the initial research concept and edited the manuscript.
%%OTS obtained the funding and edited the manuscript.
%%EBF provided the formal analysis and validation. He also edited the manuscript.
%%GEH Supervised the undergraduates, wrote the software and administers the project github and Zenodo repositories.
%%
%% Authors can use the Contributor Role Taxonomy (CRediT) at
%% for ideas on how write a good statement tailored to their needs.

\end{contribution}

\facilities{XRISM, \textit{NuSTAR}, Seimei (KOOLS-IFU), Magellan:Clay (WINERED)}

\software{
%astropy \citep{2013A&A...558A..33A,2018AJ....156..123A,2022ApJ...935..167A},  
HEASoft/FTOOLS \citep{heasoft,ftools},
IRAF \citep{1986SPIE..627..733T,1993ASPC...52..173T}, 
pyQSOfit \citep{2018ascl.soft09008G, 2019ApJS..241...34S,2024ApJ...974..153R},
WARP \citep{2024PASP..136a4504H},
TerraPCA \citep{2026ApJS..284...49M}
          }


\appendix
\section{Comparison with the Inner Radius Estimation Based on the \texttt{rdblur} Model}
\label{sec:appendix_rdblur}

In the main text, the physical radii of the torus and xBLR are constrained based on Equation~(\ref{eq:fudge}). On the other hand, the relativistic convolution model \texttt{rdblur} \citep{1989MNRAS.238..729F} is frequently employed to constrain the characteristic radius in X-ray spectral analysis \citep[e.g.,][]{2024ApJ...973L..25X,2026NatAs..10..880X}. The \texttt{rdblur} model computes the relativistic smoothing of an emission line from an accretion disk around a Schwarzschild black hole, parametrized by the emissivity index $q$ (where emissivity $\propto r^q$), inner radius $R_\mathrm{in}$, outer radius $R_\mathrm{out}$, and inclination angle $\theta_i$. When considering circumnuclear structures such as the torus or the xBLR, the emissivity index is conventionally fixed to $q = -2$ (assuming uniform illumination), and the inclination angle is fixed to an independently constrained value. Because disk emissivity declines steeply with radius, emission from larger radii contributes less to the overall line broadening, making $R_\mathrm{out}$ challenging to constrain from spectral fitting alone. Consequently, $R_\mathrm{out}$ is typically fixed at an arbitrarily large value (e.g., $R_\mathrm{out} = 10^8\ R_\mathrm{g}$ or $10^7\ R_\mathrm{g}$). Under these assumptions, $R_\mathrm{in}$ remains the sole free parameter governing relativistic line broadening. The objective of this section is to establish an empirical conversion relation between the Gaussian line width (FWHM) and the inner radius $R_\mathrm{in}$ derived from \texttt{rdblur}, thereby enabling a direct and quantitative comparison between studies employing Gaussian velocity broadening and those adopting \texttt{rdblur}.

\begin{figure}[tb]
\centering
\includegraphics[width=\columnwidth]{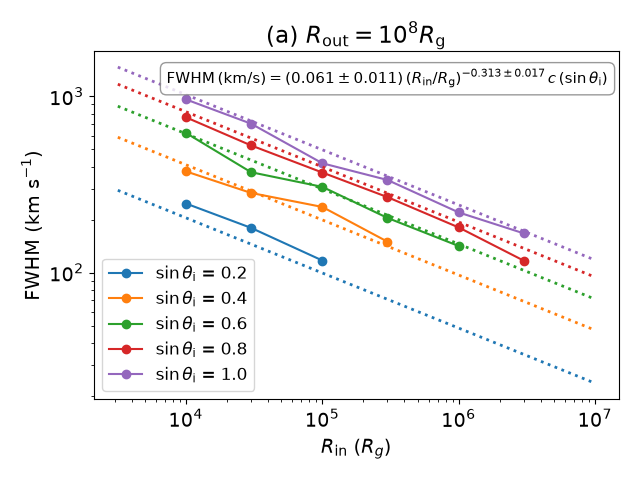}
\includegraphics[width=\columnwidth]{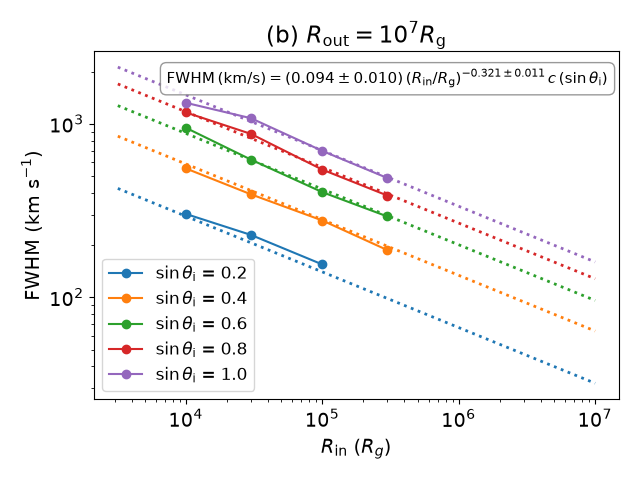}
\caption{FWHM of the \texttt{rdblur}-convolved line profile as a function of the inner radius $R_\mathrm{in}$ for various inclinations ($\sin i$), simulated with (a) $R_\mathrm{out} = 10^8\ R_\mathrm{g}$ and (b) $10^7\ R_\mathrm{g}$. The emissivity index is fixed to $q = -2$. Dashed lines represent the best-fit power-law relations.}
\label{fig:plot_sigma_q2}
\end{figure}

Assuming that the circumnuclear gas exhibits no turbulent motion, the fudge factor $f$ in Equation~(\ref{eq:fudge}) can be parametrized as $f = f^\prime \sin^{-1}\theta_\mathrm{i,torus}$. Under this formulation,
\begin{equation}
\mathrm{FWHM} = {f^\prime}^{-1} \left(\frac{R}{R_\mathrm{g}}\right)^{-0.5} c \sin \theta_\mathrm{i,torus}.
\label{eq:a1}
\end{equation}
Equation~(\ref{eq:a1}) strictly holds if the emitting region is an infinitesimally thin rotating ring. However, under the realistic assumption where the material extends continuously from $R_\mathrm{in}$ to $R_\mathrm{out}$, the power-law index of $(R/R_\mathrm{g})$ is expected to deviate from $-0.5$. Therefore, we adopt the generalized relation:
\begin{equation}
\mathrm{FWHM} = A \left(\frac{R_\mathrm{in}}{R_\mathrm{g}}\right)^{\alpha} c \sin \theta_\mathrm{i,torus},
\label{eq:a2}
\end{equation}
or equivalently,
\begin{equation}
\frac{R_\mathrm{in}}{R_\mathrm{g}} = A^{-1/\alpha} \left( \frac{\mathrm{FWHM}}{c\,\sin \theta_\mathrm{i,torus}} \right)^{1/\alpha}.
\label{eq:a3}
\end{equation}
We determine the normalization factor $A$ and the power-law index $\alpha$ through numerical simulations.

Spectral simulations were carried out using the \texttt{fakeit} command in \texttt{XSPEC}. Synthetic spectra were generated by convolving an intrinsically narrow (zero-width) Gaussian emission line with \texttt{rdblur}, and the resulting convolved profile was subsequently fitted with a standard Gaussian model to determine the velocity dispersion $\sigma_\mathrm{v}$ and the corresponding FWHM. This procedure was systematically performed over a grid of $R_\mathrm{in}$ and $\sin \theta_i$.

Figure~\ref{fig:plot_sigma_q2} illustrates the results for $R_\mathrm{out} = 10^8\ R_\mathrm{g}$ and $10^7\ R_\mathrm{g}$. The derived slopes are shallower than the theoretical value of $-0.5$. This shallower slope is attributed to the fact that, for a fixed $R_\mathrm{out}$, emission from larger radii (possessing lower orbital velocities) non-negligibly contributes to the composite line core, thereby narrowing the overall profile. This dilution effect becomes increasingly pronounced at smaller $R_\mathrm{in}$, where the dynamic range between $R_\mathrm{in}$ and $R_\mathrm{out}$ is larger. Consequently, we obtain $A = 0.061 \pm 0.011$ and $\alpha = -0.313 \pm 0.017$ for $R_\mathrm{out} = 10^8\ R_\mathrm{g}$, and $A = 0.094 \pm 0.010$ and $\alpha = -0.321 \pm 0.011$ for $R_\mathrm{out} = 10^7\ R_\mathrm{g}$.

In the hybrid spectral model presented in \S\ref{sec:hybrid}, the narrow
torus component exhibits a line width of
$\text{FWHM}=870_{-280}^{+260}$~km~s$^{-1}$.
Substituting this value into Equation~(\ref{eq:a3}) yields an inner radius
of $R_\mathrm{in} = (4.1_{-2.3}^{+10.1}) \times 10^3\ R_\mathrm{g}$ for
$R_\mathrm{out} = 10^8\ R_\mathrm{g}$, or
$(1.3_{-0.7}^{+3.0}) \times 10^4\ R_\mathrm{g}$ for
$R_\mathrm{out} = 10^7\ R_\mathrm{g}$.
These values are one to two orders of magnitude smaller than the radius of
the torus obtained in \S\ref{sec:narrow}
($R_\mathrm{torus} = (2.7_{-1.1}^{+3.2}) \times 10^5\ R_\mathrm{g}$).
Indeed, $R_\mathrm{torus}$ represents the geometric mean of the inner and outer radii.

%appendix
%% For this sample we use BibTeX plus aasjournalv7.bst to generate the

\bibliography{ms}{}
\bibliographystyle{aasjournalv7}



\end{document}